UDC 371.64:378.14
Popel Maya
**Institute of Information Technologies and Learning Tools of the National Academy of Pedagogical Sciences of Ukraine, Kyiv, Ukraine**

# THE METHODICAL ASPECTS OF THE ALGEBRA AND THE MATHEMATICAL ANALYSIS STUDY USING THE SAGEMATH CLOUD

*The quality of mathematics education depends largely on the quality of education in general. The main idea may be summarized as follows: in order to educate the younger generation of people to be able to meet adequately the demands of the time, it is necessary to create conditions for the high-quality mathematics education. Improving the quality of mathematics education of pupils in secondary school is one of the most pressing problems.*

*Contents of the school course of mathematics and its teaching method has always been the subject of undammed and sometimes stormy scientific debates. There are especially true methods of teaching algebra and the analisis in the high secondary school. Still in the study process the algebraic concepts and principles of analysis are given in such an abstract and generalized form that the student may has considerable difficulties to map these general abstract concepts to the certain concrete images, they are generalizations of.*

*Improving education quality indicators can be achieved by using the appropriate computer technology.*

*The article deals with the use of the cloud-oriented systems of computer mathematics (SCM). The prospects of development of the Web-SCM in terms of cloud-based learning environment are considered. The pedagogical features of the SageMath Cloud use as a tool for mathematics learning are revealed. The methodological aspects of algebra and elementary analysis teaching in a high profile school using the cloud-oriented the SCM SageMath Cloud are revealed.*

**Introduction.** The problems of the elaboration and deploy of the cloud-based systems in the educational process caused many debates and challenges being the subject of meticulous interest of the scientists and the educators-practitioners [10]. Which tools and technologies it is advisable to use in order to improve the learning outcomes, to exploit the potential of the just emerged ICT-based tools more fully, and the most important – to achieve better understanding and to get knowledge, but not to rice the complexity of access to relevant information? Now these questions are not a matter of the far future, it go into practice. In this regard, it is important to identify the areas of the cloud-based systems using, including the SageMath Cloud, in teaching mathematics.

Analysis of the recent research studies shown that the use of the cloud computing for the learning organization was made in the works of N. V. Morse, O. G. Kuzmynska; organization of independent work using the Yandex cloud services was reflected in the works of G. A. Aleksanyan; organization of the "virtual" Google-site facilities for teachers was studied by L. V. Rozhdestvenska.

The analysis of the prospects for SaaS technology use in education is provided in the work of the our researchers: V. Yu. Bykov, S. A. Semerikov, M. I. Zhaldak, N. V. Morse and others. The issue of such technologies implementation for education and science are the subject matter of the many domestic researchers works: V. Yu. Bykov, I. S. Voytovych, M. I. Zhaldak, N. V. Morse, S. A. Semerikov, V. P. Sergienko, N. V. Soroko, M. A. Shynenko. The features of the introduction of the cloud technology into the teachers professional practice were investigated by the foreign scientists: Justin Reich, Thomas Daccord, Alan November, Virginia A. Scott, Alec M. Bodzin, Beth Shiner Klein, Starlin Weaver and others.

**The research focus:** to conduct a theoretical analysis of the pedagogical use of the SageMath Cloud for the study of the algebra and the analysis.

**Methods:**
1) the analysis of the scientific and educational literature on the study;
2) the teaching observations and the interviews with teachers;
3) the analysis of the features of the SageMath Cloud use, as a tool of learning.

The choice of research topic is caused by the following factors:
– an objective necessity to implement the cloud technologies in the educational process;
– the idea of using the cloud resources in the classroom;
– the lack of elaborated methods with the use of the cloud resources;
– the study of the basics of mathematical analysis is quite complex and abstract material.

**Statement of the problem**

The quality of mathematics education depends largely on the quality of education in general. The main idea may be summarized as follows: in order to educate the younger generation of people to be able to meet adequately the demands of the time, it is necessary to create conditions for the high-quality mathematics education. Improving the quality of mathematics education of pupils in secondary school is one of the most pressing problems.

Contents of the school course of mathematics and its teaching method has always been the subject of undammed and sometimes stormy scientific debates. There are especially true methods of teaching algebra and the analisis in the high secondary school. Still in the study process the algebraic concepts and principles of analysis are given in such an abstract and generalized form that the student may has considerable difficulties to map these general abstract concepts to the certain concrete images, they are generalizations of.

One of the major problems of teaching methods development is the widespread adoption of the information and communication technologies (ICT) in the learning process. First of all, this problem has a significant impact on the educational system, as the use of ICT provides more opportunities in relation to processing, ordering, receiving and processing new information. So the potential and perspectives of the emerging ICT-based tools use in the field of mathematics education is a promising way of research.

**Results of the study**

The opportunities to perform the cumbersome calculations, research and visualizing the mathematical objects through the use of the mathematical software – the Systems of Computer mathematics (SCM) or the Systems of Computer Algebra are the popular subject of current research. By the definition of T. V. Kapustina, " the computer mathemativs system – is an integrated software product that combine the properties of the algebra learning systems and the universal computing environments" [5, p. 62].

SCM is equipped with the user-friendly interface, the powerful drawing tools, the significant number of tools to do math, functions and methods are implemented. Among the specific characteristics of the modern SCM the next may be noted: the availability of the inherent to these systems programming language; the tools to import the data from the other software, the tools for printing of the texts and other facilities.

"According to the research conducted by S. Stenhaus the most popular SCM are <...>: Mathematica, MATLAB, Maple, GAUSS, Scilab, you can also add MathCAD, Maxima and Sage» [5, c. 64]. The Sage – is a freely distributable software for numerical calculations and symbolic transformation and visualization of the mathematical relations and patterns in the data that is available as a web-service. "The project managed by the Sage Professor, Department of Mathematics, University of Washington (Seattle), William Stein <...>. The ultimate goal is to create an open high-quality software as a worthy alternative to the commercial software, such as Maple, Mathematica, Matlab etc.. Sage system is equipped with the two main interfaces – the local command-line interface and the Web-interface "[9, c. 81].

At present, a new class of SCM, which is designed for the network work, so-called the Web-SCM are developed. This class does not require installation of the computational kernel on your device. To begin using the system the user accesses directly to mathematical server. All computational results you get are maintained on the Web-Browser [4]. "Representatives of the Web-SCM are currently the Mathcad Application Server, MapleNet, Matlab Web Server, webMathematica, wxMaxima and Sage» [9, с. 82].

Key Features of the Web-SCM Sage are shown in table [4] (table 1):

Table 1.

*Key features of the Web-SCM Sage*

| **Advantages** | **Disadvantages** |
|---|---|
| - The openness of the system;<br>- Free distribution;<br>- A full-functional Web-server system;<br>- The integration of more than 100 mathematical packages in a single environment,<br>and others. | - The lack of scientific and methodological literature in Russian and Ukrainian languages;<br>- Not very high speed;<br>- Cumbersome interface, need much mastering;<br>- Not personalized access. |

Significant opportunities are revealed as for using Web-SCM Sage in learning mathematics [6], so it is possible:

1) to perform the calculations: both analytical and numerical;

2) to present the results of the calculations in natural, mathematical language, using symbols;

3) to build two- and three-dimensional graphs of curves and surfaces, histograms, and any other images (not excluding animations);

4) to combine computing, text and graphics in a single worksheet with the possibility of printing, publishing online and collaborate on them;

5) to create the models for practical tasks and educational research using the built-in Sage Python language;

6) to create the new classes of functions and language Python.

The SageMath Cloud – is a free service that is sponsored by the University of Washington, the National Science Foundation and Google. The SageMath Cloud was designed to facilitate the use of mathematical computing platform Android.

In The SageMath Cloud all the possibilities that are available in Web-SCM SAGE are realized, but there are some differences (table 2).

In the higher (line) school mathematics the teaching is differentiated into three levels: the level of the standard, and academic and profile. Each of these levels has both content and organizational and methodical features.

Mathematics learning at the core level is intended, above all, at the formation of the high secondary school students mathematical activities with the profound assimilation of the subject and a focus on future profession directly related to the mathematics or its applications. Training in specialized classes implies an increase in the proportion of self study learning, cognitive and practical activities. It is advisable to organize the research work in the classroom and after-school elective courses.

Table 2.

*Key features SageMath Cloud*

| Advantages | Disadvantages |
|---|---|
| - Improved user interface;<br>- Possibility of integration with other services;<br>- Instead of a notebook – an Account;<br>- The use of a document at the same time by more than 300 users;<br>- The possibility of developing a web server in Python;<br>- Increased performance in several times. | - The file is not always uploaded to the device;<br>- A sheet of the old format cannot be opened;<br>- No ability to view the public projects;<br>- Does not guarantee a full protection of user information. |

The state program on algebra and analysis at the profile level compared to the corresponding academic level program involves the study of a wide range of educational topics, as well as significantly higher demands on students' educational progress [3, p. 3-4]. Comparing the thematic plans at the different levels of mathematics learning, we conclude that the emphasis on core level had been made on the topics related to the analysis. This is due to modeling in the educational process of the elements of mathematics specialist professional training. In fact, the students enrolled in a class of mathematical, physical and physics-mathematical type, plan to link their learning precisely to mathematics. Basic concepts of the elementary analysis taught in the high secondary school are the basis for further study of higher mathematics in the high school. For these reasons, it is advisable to account for the specifics of algebra and analisys study at the core level of training. The new concepts are difficult for students due to the a high abstraction level, but extremely important when considering the future of mathematics learning at the high school.

One of the main lines of a content of the "Mathematics" course in a high secondary school is the functional line. An important conclusion of the functional line of the course is to consider the concepts of a derivative and an integral, which are necessary tool for the movement study.

The ability to carry out the numerical experiment and to perform quickly the necessary computing or image building, to check out a particular hypothesis, to test one or the other methods for solving problems, be able to analyze and explain the results obtained using a computer, finding out the limits of the ICT use or restriction of a chosen method of the problem solving are crucial for the study of mathematics [1, p. 30].

Presenting of the mathematical analysis principles including the theme "Derivative and its application", using the lecture schemas, demonstrations, images saves the considerable number of hours. Due to the high level of abstraction the theme requires the visual materials.

Using the graphic images we can acquaint the students with the abstract mathematical concepts to be difficult for understanding, that leads to activation of the students knowledge of the new material.

The main type of demonstration materials that can be used as a visual aids to increase the motivation are the so-called lecture demonstrations. " The lecture demonstration are the GUI applications with the semi-automatic control, which illustrate the theoretical concepts, theorems, methods, etc." [7]. Preferred in this case are using of the graphical primitives to compose it, so as to get a schematic drawing, illustration. In this case, the use of standard functions for working with graphics in the SageMath Cloud environment are preferable [8]. The use and study of such models is intended to gain the understanding of the mathematical, physical nature of the methods and algorithms more easily; being more aware of new material and a semantic foundation for solving applied problems and improves the cognitive activity through the visualization [8].

The lecture demonstrations involve the multiple calculations performing for the different values of the input parameters, so in the process of its elaboration it is reasonable to use the visual controls such as "input field", "slider", "checkbox", "menu" using appropriate functions of the SageMath Cloud. After making the appropriate code, the graphics data controls appear along with the results of computation in a field output. By means of the SageMath Cloud the different elemrnts can be build: points, lines, graphs of functions, circle, circle sector, bar chart, contour lines, vector field. Also it is possible to perform a variety of constructions on a space, such as a point, a broken, a sphere, a correct polyhedron.

The students independent work is one of the major means for systematic and fast learning. In its didactic purpose the independent work can be divided into the two main types: the training and the supervising (with O. I. Levus). It is advisable to organize independent work as research. Such studies may be proposed for students to be fulfilled at home or in the classroom by filling it in a form of a written work. The tasks for the partial research activity should be selected in the zone of proximal development so that the students have a free choice of tasks. Mastering the scientific content, a student not just receives new information, but also transforms it from personal experience, building a model of subjective knowledge, which includes not just logically essential, but also personally meaningful signs of the cognitive objects.

The SageMath Cloud use much simplifies the job to release time for research, to get rid of the routine calculations. The good results will be the hypotheses formulation and proof of it in the process of research.

To organize the knowledge and skills control of the students by means of the the SageMath Cloud is not just easy. First of all, due to the fact that the teacher should constantly monitor the students activities or to check the progress of their thoughts. The SageMath Cloud may be only a tool to facilitate this work.

The tasks may be organized in such a way that the student calculating some data in writing was proposed to make verification by the model and continued the work with it independently, thus obtaining new results are calculating it automatically.

Models that are created in the SageMath Cloud, involve changing the features, research options, and more. By these capabilities both the students can benefit for the self-performed work check, and also the teachers, for changing settings to get the correct answers for each of the tasks performed by the students.

Of course, the proposed model in any way cannot completely replace the traditional forms of the students self-study. It is only to create the conditions for creativive approach to learning, stimulating the interest for the further study of the topic.

The goal set at the beginning of the work has been achieved. The prospects of the cloud-oriented learning systems, including the SageMath Cloud use in the mathematics learning are revealed.

The proposed models are: the lecture demonstrations, the visual aids, the exercise simulators. These models are dynamic, providing its re-use. There are also the models to support the basic principles of the mathematical analysis relating to the theme "derivative and its applications." The models consist of the appropriate controls, such as a slider box, a cell to enter, menu and more. Each item is accompanied by a textual label control. In addition, each model includes the specific instructions that help you to learn. That is, the each program is fairly easy to use and intuitive.

Within each model there are basic theoretical information that you can use to perform the calculations by hand, to compare the results, to follow the progress of the work.

There is possible to use this research:
- In the teacher practice in general schools;
- In the student training process in the pedagogical institutions.

Guidelines:
- To use the SageMath Cloud purposely and methodologically correctly, so as to gain further activating the students activity and thereby to improve the learning outcomes;
- The SageMath Cloud can be primarily used for the students self-study to deepen knowledge, for hypothesis testing, for research and discovery of a new properties of a mathematical objects;
- To combine skillfully the traditional and innovative teaching methods with the use of the cloud technologies, realizing a new modern approach to the student learning.

There are the following conditions of the educational process organization using SageMath Cloud:

1. Submission of the curriculum should be concise, accessible and academic.
2. Use the computer only in case that the learning of a new concept require greater clarity or speeding up the pace of the lessons.
3. The SageMath Cloud use should be dosed.
4. Provide all the necessary conditions of the students working in the class. (Not allowed to use the same computer at the same time by the two students).

**The analysis and evaluation of the promising ways of development.** Improving education quality indicators can be achieved by using the appropriate computer technology. By means of the SCM the problem of taking into account the various learning rates and in particular, the depending of the student individual capacity level can be reasonably solved.

The methods of the concepts of "border", "original", "integral" studying are quite specific, as in previous classes the propaedeutics of these concepts is absent. The question arises as for the potential of the SCM use to facilitate their understanding in mathematics lessons. In this case, the use of information communication technologies is to improve the efficiency and quality of education, to enchance motivation. A computer combines the several advantages of the educational technology aids that usually used primarily to enhance visibility.

Introduction of the SageMath Cloud into the teaching practice provides a transition from the reproductive nature of the mechanical to providing the teaching and learning with the research nature. This increases the independence of students work, encourages them to acquisition and application of new knowledge [2, p. 134].

**Conclusions.** Using the cloud technology in mathematics learning is a promising way for development and improvement of the process of study. Therefore, such software tool as the SageMath Cloud, has a significant potential to improve the quality of mathematical training of the students. The prospects for further research could be the development of the SageMath Cloud use methods towards the implementation in the process of the mathematics learning in the pedagogical university.

**Попель М. В.**
**Інститут інформаційних технологій і засобів навчання Національної академії педагогічних наук України, Київ, Україна**
**МЕТОДИЧНІ АСПЕКТИ ВИКОРИСТАННЯ SAGEMATH CLOUD У НАВЧАННІ АЛГЕБРИ І ПОЧАТКІВ АНАЛІЗУ**

Якість математичної освіти багато в чому залежить від якості освіти в цілому. Основну ідею, можна резюмувати наступним чином: для того, щоб виховувати молоде покоління людей, щоб мати можливість гідно зустріти вимоги часу, необхідно створити умови для утворення високоякісної математики. Підвищення якості математичної освіти учнів в середній школі є однією з найбільш актуальних проблем.

Зміст шкільного курсу математики і її методу навчання завжди був предметом дослідження, а іноді бурхливих наукових суперечок. Є спеціальні методи навчання алгебри та аналізу у вищій школі. Проте в навчальному процесі алгебраїчні поняття та принципи аналізу наведені в такій абстрактній і узагальненій формі, що студент може має значні труднощі для зіставлення цих загальних абстрактних понять з певними конкретними образами.

Поліпшення показників якості освіти може бути досягнуто за допомогою відповідного комп'ютерних технологій.

Статтю присвячено проблемам використання хмаро орієнтованих систем комп'ютерної математики (СКМ). Розглянуто перспективи розвитку Web-СКМ в аспекті хмаро орієнтованого середовища. Виявлено педагогічні особливості застосування SageMath Cloud як засобу навчання математичних дисциплін. Розкрито методичні аспекти навчання алгебри і початків аналізу у старшій профільній школі за допомогою хмаро орієнтованої СКМ SageMath Cloud.

**Ключові слова:** хмарні технології; хмарні обчислення; СКМ; Web-СКМ; SageMath Cloud

**Попель Н. В.**
**Институт информационных технологий и средств обучения Национальной академии педагогических наук Украины, Киев, Украина**

**МЕТОДИЧЕСКИЕ АСПЕКТЫ ИСПОЛЬЗОВАНИЯ SAGEMATH CLOUD ДЛЯ ИЗУЧЕНИЯ АЛГЕБРЫ И НАЧАЛ АНАЛИЗА**

Качество математического образования во многом зависит от качества образования в целом. Основную идею, можно резюмировать следующим образом: для того, чтобы воспитывать молодое поколение людей, чтобы достойно встретить требования времени, необходимо создать условия для образования высококачественной математики. Повышения качества математического образования учащихся в средней школе является одной из самых актуальных проблем.

Содержание школьного курса математики и ее метода обучения всегда было предметом исследования, а иногда и бурных научных споров. Есть специальные методы обучения алгебры и анализа в высшей школе. Однако в учебном процессе алгебраические понятия и принципы анализа приведены в такой абстрактной и обобщенной форме, что студент может испытывает значительные трудности для сопоставления этих общих абстрактных понятий с определенными конкретными образами.

Улучшение показателей качества образования может быть достигнуто с помощью соответствующих компьютерных технологий.

Статья посвящена проблемам использования облачно ориентированных систем компьютерной математики (СКМ). Рассмотрены перспективы развития Web-СКМ в аспекте облачно ориентированной среды. Выявлены педагогические особенности применения SageMath Cloud как средства обучения математическим дисциплинам. Раскрыто методические аспекты обучения алгебре и началам анализа в старшей профильной школе с помощью облачно ориентированной СКМ SageMath Cloud.

**Ключевые слова:** облачные технологии; облачные вычисления; СКМ; Web-СКМ; SageMath Cloud